\documentclass[final,3p,times]{elsarticle}

\usepackage{amssymb}

\usepackage[T1]{fontenc}
\usepackage[cp1250]{inputenc}
\usepackage{graphicx}
\usepackage{placeins}
\usepackage{dsfont}
\usepackage{amsmath}
\usepackage{alltt}
\usepackage{epstopdf} 
\usepackage{multirow}

\begin{document}



\begin{frontmatter}
\title{Can banks default overnight? Modeling endogenous contagion on O/N interbank market}

\author[SGH,NBP]{Pawe{\l} Smaga}
\author[UW]{Mateusz Wili{\'n}ski}
\author[UW]{Piotr Ochnicki}
\author[Poznan]{Piotr Arendarski}
\author[UW]{Tomasz Gubiec}
\address[UW]{Faculty of Physics, University of Warsaw}
\address[Poznan]{The Poznan University of Economics and Business}
\address[SGH]{Warsaw School of Economics}
\address[NBP]{National Bank of Poland}

\begin{abstract}
We propose a new model of the liquidity driven banking system focusing on overnight interbank loans.
This significant branch of the interbank market is commonly neglected in the banking system modeling and systemic risk analysis.
We construct a model where banks are allowed to use both the interbank and the securities markets to manage their liquidity demand and supply as driven by prudential requirements in a volatile environment.
The network of interbank loans is dynamic and simulated every day.
We show how only the intrasystem cash fluctuations, without any external shocks, may lead to systemic defaults, what may be a symptom of the self-organized criticality of the system.
We also analyze the impact of different prudential regulations and market conditions on the interbank market resilience.
We confirm that central bank's asset purchase programs, limiting the declines in government bond prices, can successfully stabilize bank's liquidity demand.
The model can be used to analyze the interbank market impact of macroprudential tools.
\end{abstract}


\end{frontmatter}



\section*{Acknowledgements}
Support of Fundation For Polish Science (28/UD/SKILLS/2015) in funding this research is acknowledged. 
The opinions expressed herein are those of the authors and do not reflect those of the associated institutions. 
The authors are grateful to Professor Stefan Thurner from Medical University of Vienna for helpful discussion and comments.

Corresponding author's email: Tomasz.Gubiec@fuw.edu.pl.

\clearpage

\section{Introduction}

The interbank market activity in many countries has been severely impaired during the recent global financial crisis.
The events of 2007 were the hard way to find out how a single shock can lead to disastrous effects on the whole international financial system.
The network of complicated relations and dependencies between financial institutions across the globe was the main reason for which a single crash spread through the world like a disease \cite{helbing2013globally}.
From this moment on, the term \textit{contagion} \cite{stiglitz2010contagion} became an important topic in financial stability research.

The central term in mentioned area of research is \textit{systemic risk}\footnote{See \cite{de2000systemic} and \cite{smaga2014concept} for an overview of the systemic risk concept.}.
In economy and finance, risk is often described in terms of volatility and uncertainty.
A more precise description and at the same time a good measure of risk, would be to say that it is a probability of an event multiplied by its negative consequences.
What this approach lacks are the dependencies and correlations between events.
Thus, systemic risk is not about a single, unrelated crashes, it is about connected series of negative events.
In more descriptive way, it is about financial \textit{domino effect} or an avalanche of failures, where even smaller but correlated events, may lead to system breakdown.
Study of systemic risk aims at describing the odds of the whole system collapse, instead of concentrating only on individual institutions risks.
A very elegant description of the systemic risk concept can be found in Haldane et al. \cite{haldane2011systemic}.
One of the crucial elements in the rise of so called \textit{cascades} of failures are connections between different elements of the system.
These connections may transmit negative effects from one institution to another causing a great damage to the whole economy.
In order to approach this problem, we need to precisely define the links.
Having that, we will be able to use the network formalism, a common tool used by researchers in many areas \cite{fang2007new} including finance and economy\footnote{In 2013, Nature Physics published a special issue, which main subject was 'Complex Networks in Finance'. http://www.nature.com/nphys/journal/v9/n3/index.html}.
This popularity is partly because of clear analogies between financial systems and other systems where this approach turned out to be very succesful \cite{may2008complex}.

Allen and Gale \cite{allen2000financial} where the first to show that contagion risk prevails in incomplete bank network structures where small shocks lead to large effects through contagion mechanism.
Freixas et al. \cite{freixas2000systemic} further show that interbank market used for liquidity management purposes endogenously leads to a coordination failure which exposes the system to contagion in case of a single too-interconnected-to-fail bank default.
Therefore, the loosened market discipline on  the interbank market requires the central bank to act as a crisis manager.
This notion is further strengthened by Montagna and Lux \cite{montagna2014contagion}, who show that there are non-linearities in the way bank-specific shocks are propagated and structure of the various layers of interbank relations matters for contagion assessment.
Such studies show that the network structure determines the stability of the interbank market (for a contrasting view see Birch and Aste \cite{birch2014systemic}, who argue that the network structure has less importance).

The inherent instability of the network structure is exemplified by a robust-yet-fragile character of financial networks \cite{gai2010contagion}.
The network structures are resistant to shocks in general or to peripheral nodes, yet vulnerable to contagion from failure of nodes with concentrated exposures \cite{albert2000error,doyle2005robust,langfield2014mapping}.
This is confirmed by other studies \cite{boss2004network,puhr2012contagiousness,fricke2012core,craig2014interbank} which find that majority of real interbank networks have strong tiered and core-periphery structures, meaning that they have low density and the exposures distribution is concentrated in a small number of nodes that lay in the core.
Therefore, tiered networks may be inherently less stable \cite{sachs2014completeness}.

Most of interbank network analyses are country studies based on static interbank network structures, while assuming idiosyncratic or common shocks (e.g. \cite{sheldon1998interbank,furfine1999microstructure,wells2002uk,boss2004network,upper2004estimating,lubloy2005domino,degryse2004interbank,rordam2009topology,martinez2010systemic,memmel2013contagion,fourel2013systemic}).
They are surveyed by Upper \cite{upper2011simulation}, who concludes that the common finding is that contagion in core-periphery structured networks can be considered as a low probability, but high impact event.
The drawback of those studies is that they are based on a point-in-time structure of the interbank network and due to its constant evolution the resulting contagion risk should be assessed with caution.
Moreover, the studies analyzed by Upper usually include contagion effects only in one particular segment of the interbank market (the lending market), excluding other layers of interconnectedness (e.g. cross-holding of securities).
However, different studies include randomly generated distribution of possible networks and provide more robust results \cite{halaj2013assessing}.
Taking into account banks' behavioral reactions motivated by liquidity management strategies leads to liquidity contagion (e.g. \cite{karas2008liquidity,aikman2011funding,bluhm2011default}) which cannot be neglected.
We contribute to this debate by simulating a dynamic market structure with liquidity being one of the main drivers of banks' behavior in our model.

The main idea behind the network analysis is that the properties and behavior of a node have to take into account the behavior of nodes directly or indirectly linked to it \cite{ecb2010june}.
In most cases of systemic risk research, nodes in the analysed network are simply financial institutions.
In this particular paper we focus on the interbank system, thus analysed network will consider only banks.
Moreover, we want to concentrate on the transactions with short-term maturities, mainly overnight loans.
This type of loans is a substantial part of the whole interbank activity, and their main purpose is to manage liquidity.
In most countries they represent about 30\% of all interbank loans volume, making them an important and not negligible part of the market.
Furthermore, in countries like Russia, Poland, Czech Republic or Belarus, overnight loans dominate entirely, having more than 90\% share in the interbank market.
Therefore, we believe that analysing the impact of overnight loans on market stability, is an important part of systemic risk research, maybe the most important for some parts of the world.
As a result, in our model links between banks describe overnight loans.
The main obstacle in majority of networks analysis studies is lack of data on bilateral interbank exposures either from financial reporting or payment system flows.
This data is confidential and usually unavailable even for supervisory authorities and central banks.
As a solution to these problems, many researchers use various methods, in order to estimate the structure of connections between banks \cite{lu2011link,halaj2014modeling,musmeci2013bootstrapping}.

Network approach has been intensively used to analyze the interbank market contagion.
This market can on the one hand act as a shock absorber, bank liquidity management mechanism and induce market discipline through incentivizing monitoring (e.g. \cite{rochet1996interbank}).
But on the other hand, interbank market may serve as a transmission channel for contagion effect \cite{nier2007network,allen2008networks}.
There are two crucial propagation mechanisms in the interbank market.
The most obvious one is simply through insolvency of debtors, who are unable to pay off their liabilities.
In this case, negative effects spread along the edges of our network.
This transmission channel is direct.
In contrast, the \textit{trust effect} causes a financial crisis to propagates indirectly.
In response to other institutions financial problems, a bank may decide to limit its participation in the interbank market.
Such behavior leads to an overall reduction in bank funding supply on the interbank market or selectively only towards banks perceived as unhealthy \cite{arinaminpathy2012size,karas2008liquidity,de2009systemic,anand2012rollover,gabrieli2015cross}.
As shown by Gai and Kapadia \cite{gai2010liquidity}, this propagation mechanism may cause severe cascade effects.
However, interbank is not the only source of failure propagation.
Crisis may also spread due to common external assets held by different banks.
If a failing bank decides to sale its assets in order to maintain liquidity, he may cause a significant loss in value of a particular asset or even a whole class of assets, making him also suffer a loss.
If other banks, with a similar portfolio of external assets, are on the verge of solvency, it may trigger a series of defaults across the system \cite{alessandri2009framework,gauthier2010macroprudential}.
The model presented in the paper accounts for all of the above mechanisms of contagion.

The approach to modelling economical phenomena, shown in this work, is a part of an increasingly popular concept of agent based modelling in finance and economy \cite{gatti2011macroeconomics}.
Complexity theory, agent based models and network science are the elements that may give rise to a new quality in modern economy and finance \cite{farmer2009economy,battiston2016complexity}, and they are all jointly used in our research.
The structure of our model, which will be described in further sections, is in some aspects similar to \textit{model B} proposed in paper by Iori et al. \cite{iori2006systemic}.
Instead of analysing single shocks, we try to find a relation between the size of banks cash fluctuations and the stability of the market.
In such a case, collapse will be a result of the internal system dynamics.
We make several assumptions regarding the response of a bank to these fluctuations given the need to fulfill the simplified prudential requirements. 
We also need to remember that those random changes in bank's balance sheet may be both positive and negative.
Finally, even though banks react in a rational way from their individual perspective, the decisions they make often have a negative impact on the system as a whole.
Their strategy may even assume other banks failure as long as it keeps them solvent \cite{perotti2002last}.

First step of the model simulation is to build a network of bilateral interbank exposures, according to the supply and demand on the interbank system.
Since it is very unlikely to get a perfect matching across all the banks that need money and those who want to lend money, banks need to reach to the securities market next.
At every step, banks need to check the status of their debt and the regulatory requirements that they need to fulfill.
The central questions of our paper is to what extent does the interbank market stabilize the system, what is its impact on propagation of defaults and how the regulations affect both stability and contagion process.

The contributions of the paper are threefold. 
First, unlike most previous models which tested system's reaction to an external shock, in our dynamic model crashes of the entire banking system may occur as a result of an internal feature of the system (cash fluctuations). 
Secondly, our interbank network structure is dynamic and while other studies mainly analyze static network structures (see \cite{upper2011simulation} for a survey), the network structure in our model is unique i.e. simulated every day on the basis of interbank transactions from the previous day.
From this perspective, our work is a part of large branch of science which is dedicated to systems described with evolving networks \cite{yook2001weighted} (or even more popular lately, temporal networks \cite{holme2012temporal}).
Thirdly, we  take into account three types of interconnectedness channels at the same time together with an array of capital and liquidity requirements. 
We show to what extent do prudential ratios dampen or magnify contagion on the interbank market. 
We also confirm that central bank's asset purchase programs, limiting the declines in government bond prices, may successfully stabilize bank liquidity demand.

In our study, we used data from Polish interbank market.
To the best of our knowledge, there are scarcely any studies of the interbank contagion in Poland, all of which find a limited contagion risk.
Ha{\l}aj \cite{halaj2005badanie}, studying domino effects, suggests that it results only from the failure of some of the largest interbank players and the second round effects of the initial default are negligible.
The Polish central bank (NBP) study also confirms a low probability of a domino effect \cite{nbp2011july}.
Gr{\k{a}}t-Osi{\'n}ska and Pawliszyn \cite{grkat2007poziomy}, using BoF-PSS2 Payment System Simulator, come to a similar conclusion that banks hold excess liquidity for the settlement of intraday payments, while the queue management and central bank's intraday credit in the payments systems contribute to low liquidity contagion potential \cite{nbp2013july}.

Our study has several practical applications. 
It may be successfully applied to a stress testing framework to gauge the probability and impact of shocks and the resulting network resilience. 
The model may also be used to test the interbank impact of macroprudential requirements and their calibration before they are implemented. 
Moreover, if extended, the model might prove helpful in assessing the systemic importance of particular nodes to identify the super-spreaders, which is of great relevance to both central bank liquidity management for monetary policy purposes, as well as to detection of too-interconnected-to-fail institutions \cite{leon2014identifying}.
It would be particularly fruitful to combine this model with models of long-term maturity loans, which would give us a more comprehensive picture of the interbank market.

The article is structured as follows.
First, we describe the stylized facts about the interbank market in Poland and data used in the model.
Model assumptions and its construction are presented in the subsequent section.
After that we discuss the results.
In the last part we present policy implications of our research and conclusions.

\section{Stylized facts about the interbank market in Poland and data used}

We study the Polish unsecured segment of the interbank market that includes transactions between banks used for liquidity management purposes.
According to the National Bank of Poland data \cite{nbp2014rozwoj} the unsecured segment dominated in Poland in 2013 (almost 60\% of the total volume of interbank transactions), while it played a marginal role in the euro area, dominated by repos (60\%).
Polish banking sector has a structural liquidity surplus and the counterparty exposure limits between banks on interbank market are set relatively low to limit credit risk.
The interbank market is stable and O/N constituted almost all (93\%) unsecured transactions in 2013 (both in terms of volume and value) with very few transactions for longer periods.
In the euro area the term structure of interbank transactions is more balanced.
An average daily turnover of unsecured O/N transactions in Poland equaled 4.36 bilion PLN in 2013, half compared to pre-crisis period.
 
There is a clearly visible core-periphery structure of the Polish interbank network -- almost half of the unsecured interbank market turnover (both the demand and the supply sides) is concentrated in 5 banks (75\% of O/N), according to KNF\footnote{Polish Financial Supervision Authority} data \cite{knf2012rynek}.
These banks have structural liquidity surpluses and determine the liquidity of the whole market.
For payment settlement purposes banks can also use the secured overnight credit (standing facility) from the central bank or lombard credit, yet their use is negligible.
The obligatory reserve ratio is 3.5\% and banks hold marginal excess reserves at the NBP.

There are also three features of the banking system in Poland relevant to the perspective of our model approach.
First, the Polish banking sector has a large home bias.
According to the ESRB \cite{ESRB2015report}, at end-2013, holdings of sovereign debt (domestic and other Member States) equaled 41\% of total assets, while the home bias (holdings of domestic sovereign debt/holding of domestic + euro area sovereign debt) stood at 98\% in Poland.
Second, equity in the Polish banking system is of very high quality i.e. it consists almost entirely of fully loss-absorbing Tier 1 capital.
According tho KNF data \cite{knf2015report}, Tier 1 capital constituted 90.2\% of the total equity in the banking sector at the end of 2013.
Third, there is a significant maturity mismatch in the Polish banking sector for short-term maturities.
Liabilities with maturity of up to 1 year constituted 73.5\% of the total liabilities, while on the asset side this share equaled only 29\% \cite{knf2015report}.
Therefore, liquidity risk is an important determinant of banking system stability in Poland.

As a sample set in the simulation we use end-2013 data collected for 31 largest banks in Poland from their respective annual reports on a standalone basis. In our sample we have included only the banks fulfilling the following two conditions. First, the bank had to participate in the Elixir system (interbank payments clearing system in Poland). At the end of 2013 there were 46 banks active in Elixir. Second, the bank had to provide complete financial reporting data in its annual reports. We exclude the central bank (NBP) and the State Development Bank of Poland. The final list of banks used in the model simulation is included in appendix 1. Our sample, thus, covers 79.8\% of banking sector assets in Poland as of end-2013 (including assets of commercial banks, cooperative banks and branches of foreign credit institutions). This is a relatively high share, given that the interbank transactions in Poland are performed mainly within the pool of commercial banks and branches of foreign credit institutions, while numerous small cooperative banks usually use their associated banks for liquidity purposes and do not engage in interbank transactions directly. All regulatory ratios are calculated using the same end-2013 data.



\section{The model (assumptions and construction)}

In the previous section we briefly presented characteristics of the Polish interbank market. As already mentioned, in contrast to most European countries, unsecured O/N loans dominate the interbank market in Poland. For that reason, a majority of models present in the literature tested on different market structures cannot be directly applied to the Polish market. In fact, the network of loan connections on the Polish interbank market differs every single day, while most of hitherto models analyzed the dynamics of a static loan network only. This explains why creating a new liquidity-driven interbank market model was necessary.

In this section we introduce a new model of an interbank system dominated by transactions with short maturities. Our aim is to model the condition and behavior of each bank on a daily basis. Since our model focuses on the short term, we analyze banks' reactions to changes in their liquidity demand and supply, shaped by internal as well as external factors. Our approach is founded on a stylized bank balance sheet-based model with a simplified structure and is similar to those presented by e.g. \cite{krause2012interbank,manna2012externalities,montagna2013multi,steinbacher2014robustness,birch2014systemic,hausenblas2015contagion,aldasoro2015bank}. In addition, it also develops the initial work on network modeling performed by Gai and Kapadia \cite{gai2010contagion} and May and Arinaminpathy \cite{May2010systemic}. We construct a banking system network from heterogeneous banks, all of which have a different size and balance sheet structure, based on the data extracted from their financial reports.

Our model has several assumptions: 
\begin{itemize}
\item	the entire market of interbank transactions consists only of unsecured O/N loans (in other words, we exclude interbank transactions with longer maturities and secured interbank transactions)
\item	banks' behavior is motivated by the need to fulfill the liquidity demand and maintain the regulatory ratios
\item	O/N loans are granted each day and repaid the following day
\item	banks' securities portfolios consist only of liquid Polish government bonds available for sale and valued using mark-to-market accounting
\item	after a bank defaults, its counterparty banks immediately experience losses on their interbank loans equal to the amount of the exposure (we assume the Recovery Rate is zero and Loss Given Default equals 100\%)
\item	any losses incurred by the bank are directly reducing its equity
\item	troubled banks cannot  raise additional capital overnight or be bailed out (no state aid provision by public authorities and no recourse to the Lender of Last Resort from the central bank is possible)
\item	we focus on short-term reactions of banks, changes in business models are not considered
\item	we analyze the impact only on the balance sheet items and exclude any income statement adjustments 
\item	counterparty (interbank exposures) and market risks (bond prices) are not being hedged
\item	we analyze the dynamics within a closed system -- only the banks from Poland participate in the domestic interbank market
\end{itemize}

Our stylized bank balance sheet structure includes several items. They are  used as variables representing the state of each bank every day. We assume that assets may be described by the following four variables:
\begin{itemize}
\item \emph{Loans} - total amount of loans to the nonfinancial and public sector at their book value
\item \emph{Cash} - cash and balances with the central bank
\item \emph{Securities} - consisting only of liquid government bond portfolio of a single bond type; current price of a single bond is described by the variable $p$, so the mark-to-market value of securities is equal to $p$ times \emph{Securities}
\item \emph{Interbank Assets} - total amount of all interbank O/N loans granted
\end{itemize}
Liabilities of the bank may be described by only two variables, namely:
\begin{itemize}
\item \emph{Deposits} - total amount of deposits from the nonfinancial and public sector at their book value as well as any bank bonds issued that can be redeemed at notice by the bank before their maturity without loss of value
\item \emph{Interbank Liabilities} - total amount of all interbank O/N loans received
\end{itemize}
Total Assets is a sum of \emph{Loans}, \emph{Cash}, \emph{Securities} times $p$, and \emph{Interbank Assets}, while Total Liabilities is a sum of \emph{Deposits} and \emph{Interbank Liabilities}. Equity of the bank is defined as the difference between Total Assets and Total Liabilities. \emph{Loans} to the nonfinancial and public sector remain constant. \emph{Cash} and \emph{Securities} serve as liquidity buffer.

Although the fundamental aim of banks' activity is to make profit, this activity is limited by the need to fulfill numerous prudential regulations. In our model we assume several simplified prudential regulatory constraints on banking activity that banks try to maintain at all time. These ratios act, therefore, as drivers of banks' behavior and their responses to changes in their liquidity demand\footnote{We go further than Aldasoro et al. \cite{aldasoro2015bank} where bank's optimization decisions are subject only to two standard regulatory requirements: liquidity and capital adequacy ratios}. As our model focuses solely on the short term, we assume that two prudential ratios are crucial: Reserve Requirement and Liquidity Ratio. These ratios can be expressed in the model in terms of the variables defined.

Reserve requirement is a monetary policy tool. It is an obligatory level of banks reserves set aside for liquidity reasons. It is dependent on the reserve rate set by the NBP (3.5\% as of end-2013). It is calculated on the basis of nonfinancial deposits (500k EUR is subtracted from the required reserve level), excluding interbank deposits. We model it as follows:
\begin{equation}
\mbox{\textrm{Reserve Requirement}} = \frac{\mbox{\textit{Cash}}}{\mbox{\textit{Deposits}}}
\end{equation}
Liquidity Ratio represents bank's ability to service its short term obligations. It is also based on Basel III Liquidity Coverage Ratio (LCR). LCR is defined as high quality liquid assets devided by total net cash outflows over the next 30 calendar days. In our model we assume that the Liquidity Ratio is defined as:
\begin{equation}
\mbox{\textrm{Liquidity Ratio}} = \frac{(\mbox{\textit{Cash}}+p \cdot \mbox{\textit{Securities}})}{\mbox{\textit{Deposits}}}
\end{equation}
Although only a fraction of deposits in the denominator may be regarded as short-term liabilities, we can multiply both sides of the equation by this fraction without a loss of generality. As a result, the regulatory limit is lower as it is multiplied by the share of short-term deposits in total deposits (which is lower than 100\%).

When those two crucial regulatory requirements are fulfilled, bank considers three other ratios. The first one is the Leverage Ratio representing the overall level of bank's risk exposure. It has to be equal to or greater than 3\% according to Basel standards. Another one is the Capital Adequacy Ratio which represents bank's resilience and overall soundness. It is modeled on the basis of Basel methodology of risk-weighted assets (RWA) and has to equal at least 8\%. RWA includes bank's exposure to credit risk, which is the most important element of RWA for Polish banks \footnote{It constituted 87.3\% of the total bank capital requirement at the end of 2013, according to KNF data \cite{knf2015report}}. We account for market and operational risk by setting slightly higher risk weights. Risk weights for cash and other assets (i.e. government bonds) equal 0\%. The third ratio considered is the large exposure limit representing part of bank's credit risk management requirements. It is based on Article 111 of Directive 2006/48/EC and limits bank's overall exposure to interbank market. These ratios are estimated in our model as follows:
\begin{equation}
\mbox{\textrm{Leverage Ratio}} = \frac{\mbox{\textit{Equity}}}{\mbox{\textit{Total Assets}}}
\end{equation}
\begin{equation}
\mbox{\textrm{Capital Adequacy Ratio}} = \frac{\mbox{\textit{Equity}}}{\mbox{\textit{RWA}}} \quad \mbox{where} \quad \mbox{\textit{RWA}} = 90\% \cdot \mbox{\textit{Loans}} + 20\% \cdot \mbox{\textit{Interbank Assets}}
\end{equation}
\begin{equation}
\mbox{\textrm{Large Exposure Limit}} = \frac{\mbox{\textit{Interbank Assets}}}{\mbox{\textit{Equity}}} \leq 25\%
\end{equation}
The definitions of all five regulatory ratios outlined above are naturally a simplification of the real regulatory requirements and the exact values of those ratios in our model should not be directly compared with regulatory limits. Since we do not have access to granular bank financial reporting data used for supervisory purposes, we had to rely on banks' annual reports, which rarely used exact definitions of particular balance sheet items. To mitigate this problem we calculate values of four regulatory ratios (Reserve Requirement, Liquidity Ratio, CAR, Leverage Ratio) based on empirical data for end-2013 and assume that these values are actually the regulatory requirement i.e. during the simulation each bank is trying to maintain the ratios higher or equal to their initial value. This is analogical to the assumption in the DebtRank method \cite{battiston2012debtrank,Bardoscia2015debt}. For the large exposure limit we are using the actual regulatory requirement of 25\% of equity. Of course, contrary to the four other ratios, the bank is trying to maintain its Large Exposure value below this threshold.

As mentioned before, we assume that our interbank system model is driven by the fluctuations in the environment which affect banks' balance sheets. Some loans are not repaid, assets are changing their values, customers' bank transfers are changing the level of cash and deposits. Moreover, we cannot neglect banks commercial activity, for example on FX swaps market. All of these factors would have to be taken into account in our model. At the same time we want to keep our model as simple as possible. Without a doubt, all of the defined variables describing the balance sheet of a single bank should fluctuate daily. As a compromise between simplicity of the model and its accuracy, we assume that the impact of the above-mentioned factors is approximated by \emph{Cash} fluctuation only. Moreover, fluctuation of \emph{Cash} level of each bank should somehow depend on its size. As a result, we assume that every day, for the period of one day, \emph{Cash} level of each bank differs by 
\begin{equation}
(\mbox{initial Cash}) \cdot \sigma \cdot N(0,1),
\end{equation}
where N(0,1) is a random number drawn from the normal distribution. Such a change in \emph{Cash} level can be both positive and negative. Let us emphasize that on average such an assumption does not change the total amount of cash available in the whole banking system. Parameter sigma measures the amplitude of fluctuation affecting the banking system. For $\sigma = 0$ there are no changes in \emph{Cash} level, and the higher $\sigma$ the higher the \emph{Cash} fluctuations and the more unpredictable and dangerous the environment of the banking system becomes (exemplifying a bank run, for instance). It may be regarded as an analogue of a physical quantity -- the \textit{temperature} of the banking system.

So far we defined the variables describing the state of the system and external factors influencing them. However, the most important part of our model are the system's dynamics, determined by banks' activity. Since we focus on a daily time scale, we only consider actions that a real bank would be able to take within one day, which include: selling/buying of securities, interbank lending/borrowing and -- following the DebtRank model -- repurchase of bonds issued, as included in the \emph{Deposits}. This way we exclude external financial support to the bank, or changes in retail loans and deposit policies. Such actions are usually taken by the management of the bank and require some adjustment time to be reflected in balance sheet items, which does not happen on a daily basis. This limits the applicability of our model to approximately 60 working days. A detailed description of the actions that can be taken by banks in our model is provided below.
\begin{itemize}
\item Interbank lending: decreases the amount of \emph{Cash} and increases the \emph{Interbank Assets}. By lending on the interbank market, the bank makes profit (although our model does not explicitly account for that profit) so we assume that it prefers to lend the amount of money exceeding the amount of \emph{Cash} needed to fulfill the regulatory requirements, rather than to put its excess aside. From the regulatory point of view interbank lending: decreases Liquidity Ratio, decreases CAR, but increases Large Exposure ratio.
\item Interbank borrowing: increases the amount of \emph{Cash} and increases \emph{Interbank Liabilities}. As a result both Total Assets and Total Liabilities increase. It is the cheapest and the fastest way to satisfy \emph{Cash} demand needed to fulfill regulatory requirements (Reserve Requirement and Liquidity Ratio). Impact on the regulatory ratios: increases Liquidity Ratio and decreases Leverage Ratio.
\item Buying of securities: decreases the amount of \emph{Cash} and increases \emph{Securities}. The bank profits from holding securities so we assume that it prefers to buy \emph{Securities} instead of just holding excess of \emph{Cash} that cannot be lent on the interbank market and exceeds the amount needed to fulfill the regulatory requirements.
\item Selling of securities: increases the amount of \emph{Cash} and decreases \emph{Securities}. If a bank does not have enough \emph{Cash} to fulfill the Reserve Requirement, it can sell its \emph{Securities}, if it has any. In this way the bank does not change any other regulatory ratios except the Reserve ratio.
\item Repurchase of bonds issued: decreases the amount of \emph{Cash} and decreases \emph{Deposits}. This operation is the only way a bank may improve its leverage ratio in the short term. It will only be taken in case of a too low leverage ratio and only until this requirement is fulfilled.
\end{itemize}

The impact of each of the actions outlined above on particular regulatory ratios is summarized in Table \ref{tab1}. The arrows indicate improvement/worsening of the ratio, while 0 signifies no impact.

\begin{table*}[t]
\begin{center}
\begin{tabular}{|l|c|c|c|c|c|}
	\hline
	Bank Action & \begin{tabular}{@{}c@{}}Reserve \\ Ratio\end{tabular} & \begin{tabular}{@{}c@{}}Liquidity \\ Ratio\end{tabular} & \begin{tabular}{@{}c@{}}Leverage \\ Ratio\end{tabular} & \begin{tabular}{@{}c@{}}Capital \\ Adequacy Ratio\end{tabular} & \begin{tabular}{@{}c@{}}Large \\ Exposure Limit\end{tabular} \\
	\hline
	\hline
	Grant interbank loan & $\Downarrow$ & $\Downarrow$ & 0 & $\Downarrow$ & $\Uparrow$ \\
	\hline
	Receive interbank loan & $\Uparrow$ & $\Uparrow$ & $\Downarrow$ & 0 & 0 \\
	\hline
	Buy securities & $\Downarrow$ & 0 & 0 & 0 & 0 \\
	\hline
	Sell securities & $\Uparrow$ & 0 & 0 & 0 & 0 \\
	\hline
	''Repayment of'' deposits & $\Downarrow$ & $\Downarrow$ & $\Uparrow$ & 0 & 0 \\
	\hline
\end{tabular}
\end{center}
\caption{Bank actions and their impact.}
\label{tab1}
\end{table*}
Let us now analyze the net impact of all five actions considered and the need to fulfill the defined regulatory ratios, on the variables describing banks' balance sheets. For the purposes of this analysis we assume that there are no transactional costs related to trading securities. The most important observation is that in such conditions these actions cannot affect banks' Equity. Equity will decrease only if a bank experiences losses on defaulted interbank loans or is hit by a negative daily random \emph{Cash} fluctuation. Within a single day a bank cannot increase its equity in any way, but it can only manage its liquidity and adjust the amount of \emph{Cash} and \emph{Securities}. In such case we cannot apply the standard condition of banks default being Equity lower than or equal to zero. Hence, in our model the default condition had to be defined in a different way which will be described further in the text.
In order to precisely define the possible interactions and dynamics of the model we need to specify the simulation scenario of a single day in a sequential manner. 
\begin{enumerate}
\item	\textbf{Repayment of O/N loans.} At the beginning of the day all banks are repaying their yesterday's interbank liabilities. After this action it is possible that bank's \emph{Cash} level is negative. However, we assume that at this stage such situation is possible as it may use the intraday overdraft facility offered by the central bank free of charge (we do not include the impact of collateralization of the intraday overdraft). As an exception, this step is not included in the very first day of the simulation.
\item	\textbf{Cash fluctuations.} Bank transfers executed by the customers and banks' trading activity result in changes in banks' Cash level, as described and justified previously.
\item	\textbf{Check of leverage level and possibly repurchase of bonds issued.} If the current state of a bank satisfies both Reserve Requirement and Liquidity Ratio, we are dealing with \emph{Cash} surplus. In such a case, and only in such a case, the bank will use this surplus to redeem the bonds issued, provided the leverage ratio exceeds the regulatory requirement as well. By repurchasing the bonds issued, the bank lowers its Total Liabilites and Total Assets, and thus, the leverage ratio. However, the bank is limited by the requirement to satisfy Reserve Requirement and Liquidity Ratio.

\item	\textbf{Specifying the demand for Cash.} Before any interbank loan is executed we need to specify the cash supply and demand for all banks. As mentioned, if a bank satisfies both Reserve Requirement and Liquidity Ratio, we are dealing with \emph{Cash} surplus. In such a case the bank is able to allocate the remaining excess of cash (X) on the interbank market. We decided to include another way of financial contagion which we called the \textit{trust effect}. It is well-known that during a crisis the banks do not fully trust each other and are not willing to grant loans in the same amount as before. To model this phenomenon we assume that if any of the banks in the system defaulted the trust between the banks is reduced. In such a case, instead of the whole surplus \emph{Cash} (X) the banks are willing to lend only 20\% of it (X/5). This fraction was chosen arbitrarily only to show the impact of this phenomenon on system's dynamics.
Moreover, granting a loan decreases bank's Liquidity Ratio and CAR, so the final \emph{Cash} supply of a bank is limited to a value which ensures Reserve Requirement, Liquidity Ratio, CAR and Large exposure limit are satisfied. 

If the current state of a bank does not satisfy Reserve Requirement or Liquidity Ratio we are dealing with a \emph{Cash} deficit. In such a case the bank disregards other regulatory requirements and seeks to borrow the amount needed to meet the shortfall to fulfill these two ratios.
\item	\textbf{Interbank loans market.} Once we divided banks into groups of potential creditors (with Cash surplus) and debtors (with \emph{Cash} deficit) we need to specify the loans' amounts. If the total Cash supply is not equal to the total \emph{Cash} demand, the total amount of interbank loans granted is limited to the inferior of the two. In the first model we assume that loans are distributed perfectly evenly i.e. every creditor lends to every debtor. The amount is proportional both to the creditor's supply and the debtor's demand. This may be regarded as the first step of proportional fitting used by Battiston et al.\cite{arXiv:1503.00621}.

\item	\textbf{Specifying the demand for Securities.} Since the interbank lending may not satisfy the demand for \emph{Cash} on the market in full or not allocate the cash supply available, we take into account an alternative way of gaining and allocating \emph{Cash}, namely we allow the banks to buy or sell \emph{Securities}. Such an operation involves a risk of loss, therefore, the interbank market is the preferred way of acquiring and allocating \emph{Cash} on the daily time scale. Analogically to the previously-analyzed interbank loans market, we specify banks' supply and demand for \emph{Securities}. If after the interbank market transactions the current state of a bank still does not satisfy Reserve Requirement or Liquidity Ratio, it needs to sell \emph{Securities} in the amount necessary to cover the \emph{Cash} deficit. On the other hand, if a bank is dealing with a \emph{Cash} surplus even after allocating the excess on the interbank market, it uses the rest of it to buy \emph{Securities}.

\item	\textbf{Securities market.} The securities market differs from the interbank market significantly. It is not a closed market and banks are not its only participants. For the purpose of the model we assume that other participants' supply and demand are constant, which means that only supply and demand of the banks' may influence the price of \emph{Securities}. We calculate the excess demand (ED = total demand - total supply) and following \cite{cifuentes2005liquidity} assume that the price of the securities changes according to the following formula:
\begin{equation} 
\frac{dp}{p}= \eta \cdot ED \label{eq1}
\end{equation}
where parameter $\eta$ is called the \textit{market depth} and $p$ is the price of security. As we may easily notice, if $\eta = 0$ the price of security does not change regardless of demand or supply. In that case, the securities market mechanism is a simple barter between \emph{Securities} and \emph{Cash}. It may be interpreted as an intervention of the Central Bank which decides to buy or sell \emph{Securities} at a specific price. In the real market $\eta > 0$. We assume $\eta = 10^{-6}$ which means that if banks decide to sell $10\%$ of all their Securities, their price will decrease by $3\%$. This assumption may be slightly too lenient to be realistic. It should, however, be enough to show the impact of this phenomenon.
\item	\textbf{Update of the state.} Although banks operate in such a way to fulfill all the regulatory requirements, we assume that the only default condition is $\mbox{\emph{Cash}} < 0$ at the end of the day. Thus, we only consider the softest condition but at the same time the one that renders a bank inevitably bankrupt.
\end{enumerate}
Since the above-mentioned steps of the simulation are performed every day, the resulting interbank market structure is dynamic and varies daily. 
 
In conclusion, our model takes into account three possible ways of financial contagion. The first channel is the direct interconnectedness i.e. a collapsing bank not paying its obligations inducing losses on its interbank liabilities. The counterparty banks suffer losses, which worsens their financial standing by reducing their equity. The second channel we analyze is the asset price contagion \cite{bluhm2011default} (indirect interconnectedness) through government bonds. A falling bank, after suffering a drop in equity is forced to sell its securities (government bonds) in significant amounts in order to maintain regulatory ratios. This in turn results in an immediate and significant decrease in their value. In effect, the bank not only does not recover the full value of the bonds held, but also affects the market price of the bond (fire sale effect). As a result, this decreases the value of securities held by other banks thus aggravating their condition. Lastly, we also consider a reduction of availability of interbank credit due to a loss of confidence.

\section{Results}

The dynamics of the model proposed in this paper are driven by cash fluctuations. 
These random changes of banks' assets are fundamental to the whole concept, determine banks' reactions and the stability of the system.
The amplitude of these fluctuations is controlled by a single parameter $\sigma$, strongly affecting the results of a single simulation. 
The most interesting outcome of the simulation is the fraction of defaulted banks at the end of the assumed simulation period (the number of defaulted banks divided by the total number of banks in the simulation). 
Because our assumptions are valid only at the time scale of days, we assume that 3 months (60 working days) is the maximum period of applicability of the model.
Hence, the average number of defaulted banks at the end of 60 days period as a function of parameter $\sigma$ seems to be the most reliable default metric.
Further in this section we focus on the plots presenting this metric in different model versions.

\begin{figure}[ht]
	\centering
		\includegraphics[width=0.75\textwidth]{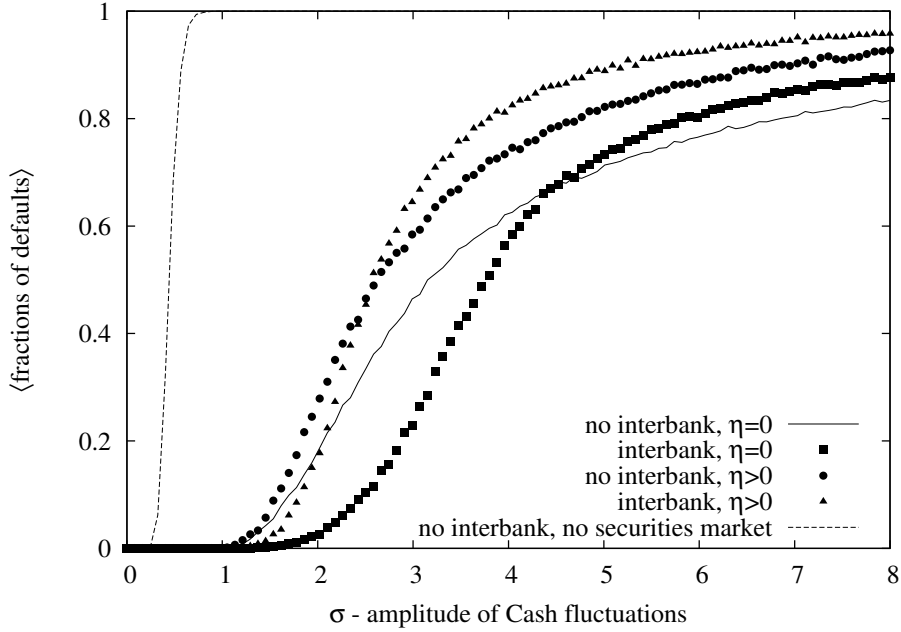}
	\caption{Average fraction of defaults as a function of $\sigma$ for various versions of the model with two regulatory ratios (Reserve Requirement and Liquidity Ratio): 
\textbf{solid black line} - $\eta=0$, without the interbank,
\textbf{black squares} - $\eta=0$, with the interbank, 
\textbf{black circles} - $\eta=10^{-6}$, without the interbank,
\textbf{black traingles} - $\eta=10^{-6}$, with the interbank,
\textbf{dashed black line} - without the securities market, without the interbank.}
	\label{fig1}
\end{figure}

A computer simulation of the banking system gives us a unique opportunity to analyze its behavior in conditions similar to the real world. We may easily exclude the selected features of the system, and thus assess their influence on the system. From this perspective, the most basic model we can think of is a case without  the securities and the interbank market. In this scenario the banks cannot react in any way, so their default is just a matter of strong enough fluctuations. Since the bank goes bankrupt when its \emph{Cash} falls below zero, the probability of default in each step is simply the probability of a negative fluctuation exceeding bank's initial cash. The distribution of the changes described is Gaussian with standard deviation equal to the initial cash multiplied by the parameter $\sigma$. Therefore, the probability of default is every day equal to $\Phi\left(-\frac{1}{\sigma}\right)$, where $\Phi(.)$ is the Gaussian cumulative distribution function. Furthermore, the probability of bankruptcy during the time period $T = 60$ days equals:
\begin{equation}
p_T(\sigma) = 1 - \Phi^T\left(-1/\sigma\right).
\end{equation}
This leads to a formula for the expected fraction of banks which went bankrupt, as a function of $\sigma$:
\begin{equation}
\langle f_{\mbox{def}}(\sigma) \rangle = 1 - \Phi^T\left(-1/\sigma\right),
\end{equation}
where $\langle . \rangle$ is the ensemble average of the process.
This result is presented in the Fig. \ref{fig1} with a dashed line. We analyzed the result for $\sigma$ from 0 to 8. Even for relatively small values of sigma, below unity, almost all banks in the system defaulted. As expected, without the interbank market and any possibility to liquidate the assets, even a relatively small fluctuation may result in a default. In this version of the model the banks are fully independent, but despite that the whole system collapses very quickly.

The obvious next step is to allow the banks to sell their securities and analyze the difference compared to the previous case.
For simplicity reasons we assume that the securities market is perfect i.e. selling or buying any amount of securities does not affect their price.
In our model such situation is equivalent to $\eta=0$ in Eq.(\ref{eq1}).
Such an assumption reflects the government bond purchase program by the central bank, being possibly one of the measures of last resort to stabilize the interbank and government bond markets during the crisis.
In this version of the model the banks cannot influence other regulatory ratios than Reserve Requirement so it is the only ratio they are trying to satisfy. This version is also analytically solvable.
The average fraction of defaulted banks as a function of $\sigma$ is given by:
\begin{equation}
\langle f_{\mbox{def}}(\sigma) \rangle = \frac{1}{N} \sum_{i=1}^{N} \left(1 - \Phi^T \left(-\frac{1+\mbox{\textit{Securities}}_i/\mbox{\textit{InitialCash}}_i}{\sigma}\right) \right),
\end{equation}
where $N$ is a number of banks in the system.
This dependance is presented in the Fig. \ref{fig1} with a solid line. 
When selling of securities is allowed, the behavior of the system is considerably different compared to the previous case.
A significant fraction of defaults is observed only for $\sigma > 1$ and even for huge fluctuations ($\sigma \approx 8$) only about 80\% of the banks defaulted.
Even such a simple example shows that purchase of government bonds in quantities large enough to stabilize their prices (\emph{quantitative easing}) may significantly decrease the number of defaulted banks during the crisis or even prevent the defaults.
In this model banks are still independent, so let us now focus on the model with interactions leading to inter-dependencies.

The most natural way of interaction between the banks is via the interbank loans market.
Let us consider the previous version of the model with perfect securities market ($\eta = 0$), but with the interbank market as described in the previous section.
In such a version, the banks are able to adjust both Reserve Requirement and Liquidity Ratio.
We, therefore, assume that banks are trying to satisfy both of these conditions.
Due to more complicated dynamics this version had to be solved using a numerical simulation. The dependance between the average fraction of defaulted bank and $\sigma$, obtained by averaging over 100 realizations, is presented in Fig. \ref{fig1} with black squares.
The result fully confirms the expectations presented in the literature (e.g. \cite{Acemoglu2015systemic,Elliott2014financial,Georg2013effect}), when compared to the version without the interbank market.
First of all, with the interbank market the number of defaults became noticable for $\sigma > 1.5$, which means that the system can survive without any defaults even for relatively large fluctuations.
Moreover, for $\sigma < 4.5$ the fraction of defaulted banks is lower than in the case without the interbank market.
However, for $\sigma > 4.5$ the fraction of the defaulted banks becomes grater than in the case without the interbank market.
This result confirms the stabilizing role of the interbank market in a situations of moderate fluctuation, but not in the case of a systemic crisis.
During the crisis, the interbank market is one of the contangion channels and may actually increase the number of defaults.
Our result confirms for instance the intuitions of the Iori et al. \cite{Iori2003interbank} not shown by their model. 

Two versions of the model presented above assumed a perfect securities market corresponding to $\eta = 0$.
Such assumption does not, however, describe the real banking system without central bank's intervention.
The next question we pondered over was the influence of a positive $\eta$.
Since we do not have the data about transactions on Polish government bonds market, we were unable to estimate the price impact function in this case and the market depth.
We limited our analysis to the qualitative study of the impact of an infinitesimally small $\eta = 10^{-6}$.
Both versions with and without the interbank market were analyzed with positive $\eta$.
The results obtained are shown in Fig. \ref{fig1} with triangles and circles respectively.
With a positive $\eta$ a single bank with a \emph{Cash} deficit selling securities may -- via fire sales -- decrease the value of the bonds held by other banks and hence affect their financial standing.
The securities market in this form may be viewed as a second channel of contagion in our model.
As we may see in Fig. \ref{fig1} in both cases higher $\eta$ increased the fraction of defaulted banks.
The influence of the interbank market is qualitatively similar to the case of $\eta = 0$ i.e. for smaller $\sigma$ it stabilizes the system and reduces the number of defaults, while for larger $\sigma$ it amplifies the contagion.
Yet, the stabilizing effect of the interbank market is it this case much weaker and is suppressed already for $\sigma \approx 2.5$.
The influence of $\eta > 0$ is, therefore, much more significant than the existence of the interbank market in our model.

\begin{figure}[ht]
	\centering
		\includegraphics[width=0.75\textwidth]{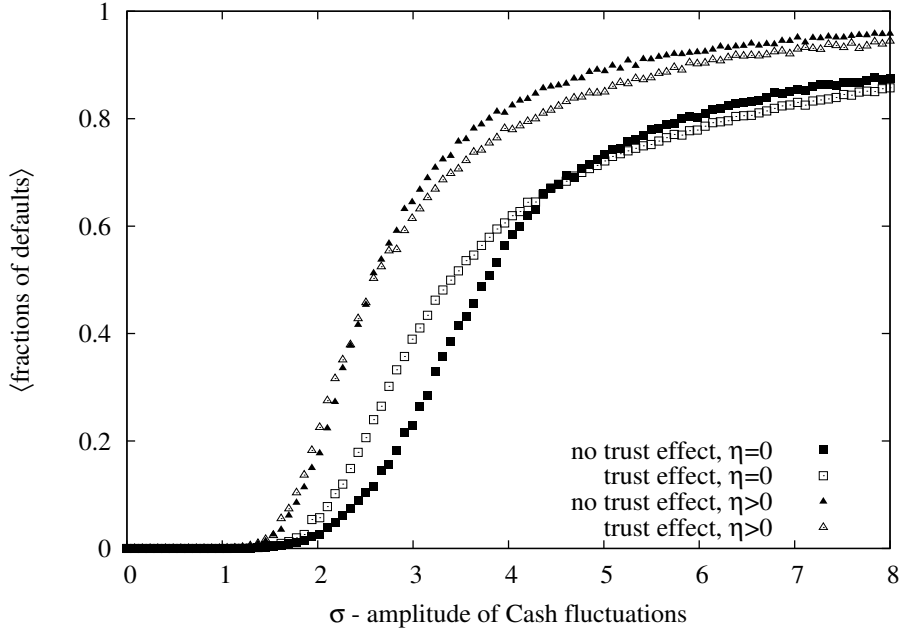}
	\caption{Impact of the \textit{trust effect}. Average fraction of defaults as a function of $\sigma$ for various versions of the model with interbank market and two regulatory ratios (Reserve Requirement and Liquidity Ratio): 
\textbf{black squares} - $\eta=0$, without the \textit{trust effect}, 
\textbf{white squares} - $\eta=0$, with the \textit{trust effect}, 
\textbf{black traingles} - $\eta=10^{-6}$, without the \textit{trust effect}, 
\textbf{white traingles} - $\eta=10^{-6}$, with the \textit{trust effect}.}
\label{fig2}
\end{figure}

In addition to two contagion channels already introduced, interbank market and securities market, we now add the third one called the \textit{trust effect}, as described in the previous section.
The comparison of the fraction of defaulted banks for various $\sigma$, with present interbank market and two cases of $\eta$ equal to zero and greater than zero are shown in  Fig. \ref{fig2}.
In both cases, $\eta = 0$ and $\eta > 0$, the influence of the trust effect is not as significant as the alterations discussed previously.
The defaults start approximately at the same point $\sigma \approx 1.5$.
For smaller sigmas the trust effect increases the fraction of defaults, but it decreases it for larger ones.
This confirms the conjecture that a decrease in interbank trust may propagate the crisis.
However, this effect is of little relevance in case of $\eta > 0$.
In the case of $\eta = 0$, which can be viewed as an intervention of the central bank, the negative impact of the \textit{trust effect} is much more explicit.
For large fluctuations the behavior described by the \textit{trust effect} is reasonable not only from the point of view of a single bank but also from the point of view of the whole system, and it slightly reduces the fraction of defaulted banks.
And while this effect significantly reduces the total volume of the interbank loans, it does not influence the fraction of defaults in a similar way.

\begin{figure}[ht]
	\centering
		\includegraphics[width=0.75\textwidth]{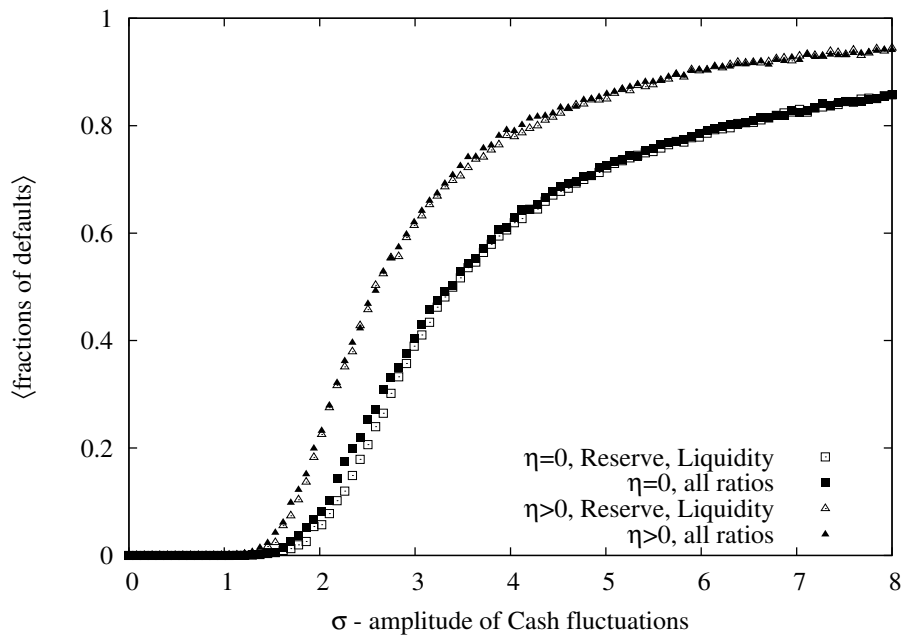}
	\caption{Impact of the regulatory ratios. Average fraction of defaults as a function of $\sigma$ for various versions of the model with interbank market and ''the trust effect'':
\textbf{white squares} - two regulatory ratios: Reserve Requirement and Liquidity Ratio, $\eta=0$, 
\textbf{black squares} - all analyzed regulatory requirements (Reserve Requirement, Liquidity Ratio, Capital Adequacy Ratio, Leverage Ratio, Large Exposure Limit), $\eta=0$,
\textbf{white triangles} - two regulatory ratios: Reserve Requirement and Liquidity Ratio, $\eta=10^{-6}$, 
\textbf{black triangles} - all analyzed regulatory requirements (Reserve Requirement, Liquidity Ratio, Capital Adequacy Ratio, Leverage Ratio, Large Exposure Limit), $\eta=10^{-6}$.}
	\label{fig3}
\end{figure}

The final inquiry made by this analysis is the impact at the daily time scale of all the remaining regulatory requirements.
Besides Reserve Requirement and Liquidity Ratio already applied, we described Capital Adequacy Ratio, Leverage Ratio, and Large Exposures limit.
The result of the simulation including all of them is presented in Fig. \ref{fig3}.
There is almost no difference compared to the previous version of the model with the trust effect, the version closest to the real banking system both in case of $\eta = 0$ and $\eta > 0$.
Upon closer inspection we are able to see two effects, both of them increasing the number of deafults.
First of them, present in the case of $\eta = 0$, increases the fraction of defaults for $\sigma \in [1.5,3]$ and is caused by the Large Exposure Limit\footnote{We analyzed all combinations of the regulatory requirements (Capital Adequacy Ratio, Leverage Ratio, Large Exposure Limit) added to the standard two (Reserve Requirement and Liquidity Ratio).
In all cases with Large Exposure Limit the effect was present.
Without the Large Exposure Limit the results overlapped the result without additional requirements.}.
Second effect infinitesimally increases the fraction of defaults for large \emph{Cash} fluctuations, $\sigma > 3$.
The latter is caused by the Leverage Ratio Requirement\footnote{The result confirmed in the same way as in the case of the first effect.}.
We may, therefore, conclude that currently used regulatory requirements have a very limited impact on the fraction of defaults for any $\sigma$ and their influence is almost negligible.

\section{Conclusions and policy implications}

The collapse of the interbank markets during the recent global financial crisis underlined the importance of the interbank contagion due to network interconnectedness.
We devised an interbank market model for an interbank market structure dominated by O/N transactions and tested it on a sample of Polish banking system data for end-2013.
We demonstrated how does banks' behavior, motivated by the need to fulfill the regulatory prudential requirements, determine the extent of contagion.
We prove that internal model dynamics may lead to contagious default cascades.
In the numerous versions of our model, we studied three possible channels of financial contagion (balance sheet interconnectedness, asset price contagion and confidence decline).
Furthermore, our results also point to some key policy implications.

Firstly, and rather unsurprisingly, we confirm the stabilizing effect of the interbank market, although it seems to be the weakest when all of the regulatory ratios considered need to be met by banks.
This underlines the conclusion that during a systemic crisis less restrictive prudential requirements may reduce the default propensity on the interbank market by maintaining its effective functioning.
However, for significant fluctuations of cash, the interbank market begins to amplify instead of dampening the shocks, which also confirms the previous findings showing that the impact of the network connectivity on the banking system's stability, depends on the  magnitude of the shock (see \cite{steinbacher2014robustness} and \cite{nier2007network}).

Secondly, once the default cascade starts, the contagion spreads relatively quickly on the market, so interventions of public authorities have to be prompt in order to be effective.
We also confirm the substitutive role of the interbank market and the government bond market as banks' liquidity demand stabilizers.
Moreover, we confirm that central bank's asset purchase programs, aimed at limiting the declines in government bond prices during the crisis, can successfully stabilize banks' liquidity demand and act as a crisis-management tool. 

Thirdly, in case of central bank's asset purchase program, the need to fulfill the large exposure limit reduces the volume of interbank transactions and slightly increases the number of defaults.
It suggests that usage of such a tool should be preceded by in-depth analyzes of applicable regulatory requirements.
In conclusion, the supervisory authorities should be able to exercise discretion and flexibly adjust these requirements, as the crisis unfolds.


We are fully aware of the limitations of our model, relating mainly to model assumptions, simplifications and assumed behavioral responses on the need to fulfill the regulatory ratios.
Therefore, our model may surely be further enhanced in many different ways e.g. by introducing the central bank as an active interbank market player or more realistic network of interbank loans connections.
Moreover, it is possible to apply our model to data for different periods and from other EU countries.
Our model may also be directly applied to data of countries with similar maturity structure of the interbank transactions e.g. the Czech Republic, Russia or Belarus, which would allow comparisons of potential contagion in other banking systems, as well as over time. Similarily, it is possible to introduce additional (i.e. regulatory or idiosyncratic) factors determining banks liquidity demand.

The results of our study contribute to the literature on the impact and effectiveness of prudential requirements on the interbank market, which is still scarce\footnote{See e.g. \cite{IMF2012theInteraction} or \cite{Akinci2015HowEffective} for a comprehensive evaluation of the effectiveness of macroprudential tools, yet not including their impact on the interbank market.}.
Our short-term model underlines the importance of studying contagion mechanisms within the daily dynamics of the interbank network structure.
This stands in contrast to the fact that a majority of hitherto models analyze contagion only over longer periods.
We believe that combining these two approaches might prove conducive to comprehensive and more precise estimations of contagion potential in the banking systems, which studies omitting the O/N transactions dynamics could underestimate.

\appendix
\section{The algorithm}
Below we describe the algorithm used in our simulation.
In order to fully understand it, it is crucial for the reader to be aware of the exact structure of the objects used in the pseudo-code.
These objects are defined as follows:
Banks -- elements of the interbank system, each characterized by the following attributes: Cash, Deposits, Equity, Loans, Securities, representing its assets and liabilities, and ReserveRatio, LeverageRatio, CapitalAdequacyRatio, LiquidityRatio, which represent the regulatory ratios imposed on each bank individually.

\begin{verbatim}
0.Setting regulatory ratios according to current Bank ratios.
    FOR EACH bank IN banks DO:
           ReserveRatio = Cash/Deposits
           LeverageRatio = Equity/( Securities + Loans)
           CapitalAdequacyRatio = Equity/Loans
           LiquidityRatio = Securities/Deposits
1.Repayment of all O/N loans.
    FOR EACH loan IN loans DO:
           IF(DebtorDefaulted = TRUE):
               Creditor[Cash] = Creditor[Cash]  + Amount
               Debtor[Cash] = Debtor[Cash] - Amount
2.Cash fluctuations.
    FOR EACH bank IN banks DO:
           Cash = Cash - RandomCash[previous_day]
           RandomCash[current_day] = InitialCash*RandomGauss(mean=0, variance=Sigma)
           Cash = Cash + RandomCash[current_day]
3.Checking leverage ratio of every bank and repaying deposits if necessary.
    FOR EACH bank IN banks DO:
           IF( Equity / ( Loans + InterbankAssets + Securities) < LeverageRatio ):
               lackLR = Loans + InterbankAssets + Securities - Equity / LeverageRatio
               maxLR = MAX((Securities - LiquidityRatio*Deposits )/(1 - LiquidityRatio),0)
               maxRR = MAX((Cash - ReserveRatio*Deposits)/(1 - ReserveRatio),0)
               Deposits = Deposits - MIN(lackLR,maxLR,maxRR)
               Cash = Cash - MIN(lackLR,maxLR,maxRR)
4.Calculating demand for cash for each bank.
    FOR EACH bank IN banks DO:
           CashDemand = 0.0
           IF(Securities/ Deposits >= LiquidityRatio AND Cash >= ReserveRatio*Deposits):
               maxLR = MAX(Securities - LiquidityRatio*Deposits, 0)
               maxBE = MAX(0.25*Equity, 0)
               maxCAR = MAX(5.0*(Equity/CapitalAdequacyRatio - Loans), 0)
               CashDemand = -TrustRatio*(Cash - ReserveRatio*Deposits)
               IF(LiquidityRatioOn = TRUE):
                      IF(CashDemand < -maxLR):
                          CashDemand = -maxLR
               IF(BigExposureOn = TRUE):
                      IF(CashDemand < -maxBE):
                          CashDemand = -maxBE
               IF(CapitalAdequacyRatioOn = TRUE):
                      IF(CashDemand < -maxCAR):
                          CashDemand = -maxCAR
           ELSE:
               CashDemand = MAX(LiquidityRatio*Deposits - Securities,
							                  ReserveRatio*Deposits - Cash,0)
5.Network of interbank loans is created according to banks' demand and supply.
 All of the loans are executed.
    Total = MIN(SUM(Supply), SUM(Demand))
    Sup = Supply/SUM(Supply)
    Dem = Demand/SUM(Demand)
    FOR i = 1 TO SIZE(Supply) DO:
           FOR j =1 TO SIZE(Demand) DO:
               Amount = Total*Sup*Dem
               IF(Amount > 0):
                      ADD Loan(Banks[i], Banks[j], Amount) TO Loans
                      Supply[i] = Supply[i] + Amount
                      Demand[j] = Demand[j] - Amount
6.Calculating demand for securities for each bank.
    FOR EACH bank IN banks DO:
           ExpectedCash = ReserveRatio*Deposits
           SecuritiesDemand = MAX((Cash - ExpectedCash)/SecuritiesPrice,-Securities)
7.Orders for securities are executed at a price calculated according to the demand size.
    TotalDemand = SUM([SecuritiesDemand FOR ALL banks]) 
    SecuritiesPrice = SecuritiesPrice*(1 + Eta*TotalDemand)
    FOR EACH bank IN banks DO:
           Securities = Securities + SecuritiesDemand
           Cash = Cash - SecuritiesDemand*SecuritiesPrice
           SecuritiesDemand = 0
8.Every bank with cash < 0 defaults and cannot repay its loans nor participate in the system.
9.Repeat from step 1.
\end{verbatim}

\bibliographystyle{elsarticle-harv} 
\bibliography{bibliografia}

\end{document}